\documentclass[10pt,a4paper]{article}
\usepackage{latexsym,epsfig,multicol}
\usepackage{color}
\usepackage{graphicx}
\usepackage{verbatim}

\def\sk{\smallskip}

\def\beq{\begin{eqnarray}}
\def\eq{\end{eqnarray}}
\def\nl{\noindent}

\def\lra{\longrightarrow}

\def\beq{\begin{eqnarray}}
\def\eq{\end{eqnarray}}
\def\beqn{\begin{eqnarray*}}
\def\eqn{\end{eqnarray*}}
\def\nl{\noindent}

\def\ed{\end{document}}


\usepackage{amssymb}
\usepackage{amsfonts}
\usepackage{amsmath}%
\usepackage{graphicx}
\providecommand{\U}[1]{\protect\rule{.1in}{.1in}}

\title{Charged Bilepton Pair Production at LHC Including Exotic Quark Contribution}
\author{E. Ramirez Barreto, Y. A. Coutinho \\
Instituto de F\'isica, \\
Universidade Federal do Rio de Janeiro,\\
Rio de Janeiro, RJ, Brazil\\
\\
J. S\'a Borges\\
Universidade do Estado  do Rio de Janeiro, \\
Rio de Janeiro, RJ, Brazil }

\date{}

\begin{document}

\maketitle

\begin{abstract}
The production of \, $W^+ W^-$ pair in hadron colliders was calculated up to loop corrections  by some  authors in the 
Electroweak standard model (SM) framework.
This production was also calculated, at the tree level, in some extensions of the SM such as the vector singlet, the fermion mirror fermion and the vector doublet models by considering  the contributions of new neutral gauge bosons and exotic fermions.  The obtained results for $e^+ e^-$ and $pp$ collisions pointed out that the new physics contributions are quite important. This motivates us to calculate the production of a more massive charged gauge boson predicted by the ${SU (3)_C \times SU (3)_L \times  U (1)_X }$ model (3-3-1 model).
Thus,  the aim of the present paper is to analyze the role played by  of the  extra gauge boson ${Z^\prime}$ and of the exotic quarks, predicted in the minimal version of the 3-3-1 model, by considering  the inclusive production of a pair of bileptons ($V^\pm$) in  the reaction $p + p \longrightarrow V^+ + V^- + X$,  at the Large Hadron Collider (LHC) energies. 

Our results show that the correct energy behavior of the elementary cross section follows from the  balance between  the contributions of the extra neutral gauge boson with those from the exotic quarks. The extra neutral gauge boson induces flavor-changing neutral currents (FCNC) at tree level, and we have introduced the ordinary quark mixing matrices for the model  when the first family transforms differently to the other two with respect to $SU(3)_L$. We obtain a huge number of heavy bilepton pairs produced for two different values of the center of mass energy of the LHC. 

\vskip 1cm 
\par

\par

email: elmer@if.ufrj.br, yara@if.ufrj.br, saborges@uerj.br

\end{abstract}
\newpage

\section{Introduction}
\par

The gauge sector of the Standard Model (SM), has 
been extensively tested by LEP, SLD and Tevatron experiments. Among these tests, the production of $W$ pairs is important because it is quite sensitive to a balance between  $s-$ and $t-$channel contributions. 
The gauge boson pair production can also reveal the nature of the triple
gauge coupling, but until now there is no evidence of the existence of anomalous
gauge couplings.
In fact, the analysis of the $Z$ transverse momentum distribution in the process $p + \bar p \longrightarrow W + Z + X \longrightarrow \ell ^{\prime} + \nu_{\ell} + \ell + \bar \ell$ ($\ell$ and $\ell ^{\prime}$ are electrons and muons) at Tevatron  ($\sqrt s = 1.96$ GeV) gives a more restrict
limit on the $WWZ$ coupling parameters  \cite{ABA}. They are: $ -0.17 \leq \lambda_Z \leq 0.21$  ($\Delta\kappa_Z =0$) and  $-0.12 \leq \Delta\kappa_Z \leq  0.29$ ($\lambda_Z=0$) assuming that $\Delta g_1^Z=\Delta\kappa_Z$. The effective Lagrangian with the parametrization
of anomalous couplings involving $WW\gamma$ and $WWZ$ is found in \cite{ZEP, HAG}.

The standard model  $W^+ W^-$, $Z^0 Z^0$ production  in $e^+ e^-$ and in hadron colliders was studied in \cite{RON, BRM, PHI, EIC}, for example; the authors have shown that the $s-$ and $t-$channels balance in $W^+ W^-$ production is essential for the good behavior of total cross sections. Such behavior must be preserved when the c.m. energy of colliding particles increase ($\sqrt s \gg M_Z$) probably producing new particles, which appear in many extensions the SM or alternative models \cite{LIT, LRM, E6M, RIZ, GEO}. For example, the $W^+ W^-$ production in linear and hadron colliders was analyzed in some extensions of the SM which have  the same gauge boson content but where the fundamental matter representation includes new exotic fermions (very massive leptons and heavy quarks) \cite{VDM}; using  the unitarity constraint the authors determined some relations between model parameters  \cite{YWW}. In the same context, the new neutral gauge boson contribution for  left-right models in $e^+ e^-$ collisions was analyzed \cite{MAA}.

At high energy, the production of ordinary or exotic gauge boson pair can be studied from  alternative models with a large particle {\it spectrum}. The boson pair production takes place  through standard and new gauge bosons  $s$-channel contribution and from $t$-channel exchange of ordinary or new fermions. For these models the number of triple gauge couplings increase and high energy processes can also reveal anomalous couplings, excluded in a previous analysis \cite{ANO}.

We are exploring the phenomenological aspects of an alternative to the SM based on the $SU(3)_c \times SU(3)_L \times U(1)_X$ gauge symmetry (3-3-1 model) \cite{PIV, FRA, RHN, TON} which predicts  new very massive particles mixed to the observed states. Together with the  exotic fermion content and an extra neutral gauge boson, the model includes gauge bosons  carrying lepton number equal 2, called bilepton \cite{CUY}, that also occur in $SU(15)$ grand unified theories \cite{SU15}. There exist, in the literature, the supersymetric extension of the model \cite{MAR} and several phenomenological consequences of the model are being explored \cite{PIR}.

Let us outline some features of the model considered in this paper. Although at low energies the model coincides with the SM, it offers an explanation for basic questions such as the family replication problem and the observed bound  for the Weinberg angle \cite{VIC}. The family  problem is solved by considering the model  anomaly cancellation procedure, requiring that the number of fermion families must be a multiple of the quark color number  \cite{PIS}. Considering that QCD asymptotic freedom condition is valid only if the number of families of quarks is  less than five, one concludes that there are three generations.  On the other hand, to keep the validity of perturbation calculation,  one obtains  a bound for the Weinberg angle at each energy scale $\mu$, ($\sin^2 \theta_W (\mu) \le 1/4$) this constraint follows from the coupling constants ($g_{U(1)}$, $g_{SU_L(3)}$) ratio. The experimental value of $\sin^2 \theta_W (M_Z) \simeq 1/4$  leads to an upper bound associated with the spontaneous $SU(3)_L$ symmetry breaking \cite{ALD,ALE}, which implies directly on a restriction on exotic boson masses \cite{NGL}.

Working with two versions of the 3-3-1 model,  we have analyzed the $e^+ + e^- \lra f + \bar f$ , (where $f$ denotes ordinary leptons or quarks) for ILC energies in order to establish some signatures of the extra neutral gauge boson  ${Z^\prime}$ existence and to  obtain lower bounds on its mass. The obtained bounds were confirmed by extending our analysis to $p \bar p$ and $p p$ collisions \cite{EUR}. 
In another publication, we  have included the  ${Z^\prime}$  contribution to the production of a pair of double charged bileptons in $e^+e^-$ for ILC energy \cite{PLB}. The  beginning  of activities of the LHC, operating  at high energy, opens the search for new discoveries. Among these findings one expect the presence of some signatures for new particles as those predicted by the 3-3-1 model, in particular new gauge bosons, bileptons and exotic quarks. 

In the present paper we analyze the production of a pair of single charged bilepton ($V^\pm$) at the Large Hadron Collider (LHC) at CERN with  $\sqrt s > 10$  TeV, through  the process $p  + p \lra V^+ + V^- + X $,  where $s-$ channel contributions come from $\gamma$, $Z$ and $Z^\prime$  and  where $t-$ channel includes only the exotic quarks contributions.  Our calculation is performed at the tree level employing parton distribution functions \cite{CTE} in a Monte Carlo code. 

In the section II we review the basic aspects of the minimal version of the 3-3-1 model. In the section III we present the calculation of $q + \bar q \lra V^+ + V^-$  cross section as well as the final results for $p + p \lra V^+ + V^- + X$ adding some comments of our results.  Finally, in the section IV, we present the conclusions of our work.

\section{Model}

In the 3-3-1 model the electric charge operator is defined  as:  
\begin{equation}
Q = T_3 + \beta T_8 + X I
\label {beta} 
\end{equation}
\noindent where $T_3$ and $ T_8$ are two of the eight generators satisfying the $SU(3)$ algebra
\begin{equation}
 \left[ T_i\, , T_j\, \right] = i f_{i,j,k} T_k \quad i,j,k =1 .. 8,
\end{equation}
\noindent  $I$ is the unit matrix and $X$ denotes the $U(1)$ charge.

The electric charge operator determines how the fields are arranged in each representation and depends on the $\beta$ parameter.  
Among the possible choices, $\beta = -\sqrt 3$  \cite{PIV,FRA} corresponds to the minimal version of the model that is used in the present application. 

The lepton content of each generation ($a = 1, 2,  3$) is: 
\begin{eqnarray}
\psi_{a L} = \left( \nu_{a} \ \ell_a \  \ell^{c}_a \right)_{L}^T\ \sim\left({\bf 1}, {\bf 3}, 0 \right), 
\end{eqnarray} 

\noindent where $\ell^c_a$ is the charge conjugate of $\ell_a$ ($e$, $\mu$, $\tau$) field. Here the values in the parentheses denote quantum numbers relative to $SU(3)_C$, $SU(3)_L$ and $U(1)_X$ transformations.  

In order to cancel anomalies, the first quark family is accommodated in $SU(3)_L$ triplet and the second and third families ($m =2,3$) belong to the  anti-triplet representation, as follows:
\begin{eqnarray}
&& Q_{1 L} =  \left( u_1 \ d_1 \  J_1
\right)_{L}^T \ \sim \left({\bf 3}, {\bf 3}, 2/3 \right), \nonumber \\
&& Q_{m L} =  \left( d_m \  u_m \ j_m
\right)_{L}^T \ \sim \left({\bf 3}, {\bf 3^*}, -1/3 \right). 
\end{eqnarray}
\begin{eqnarray}
&& u_{a R}\ \sim \left({\bf 3}, {\bf 1}, 2/3 \right), \  d_{a R} \ \sim \left({\bf 3}, {\bf 1}, -1/3 \right),\nonumber \\
 && J_{ 1 R}\ \sim \left({\bf 3}, {\bf 1}, 5/3 \right), \  j_{m R} \ \sim \left({\bf 3}, {\bf 1}, -4/3 \right),
\end{eqnarray} 

\noindent where $a = 1,2,3$ and $J_1$, $j_2$ and $j_3$ are exotic quarks with respectively $5/3$, $-4/3$ and $-4/3$   units of the positron charge ($e$).

This version  has five additional gauge bosons beyond the SM ones. They are: a neutral {$Z^\prime$} and four heavy charged bileptons, ${Y^{\pm\pm},V^\pm} $ with lepton number {$L = \mp 2$}.  
In order to avoid model anomalies, only one quark family must be assigned to a different $SU(3)$ representation, but this procedure does not specify what is the family to be elected \cite{CAR}.  We will comment, in the conclusion section,  about the consequences of our choice where the first family is treated differently from the other two.  

The minimum Higgs structure  necessary for symmetry breaking and
that gives quark and lepton acceptable masses are composed by three triplets  ($\chi$, $\rho$, $\eta$) and one anti-sextet ($S$). 
The neutral field  of each scalar triplet develops non zero vacuum expectation values ($v_\chi$, $v_\rho$, $v_\eta$, and $v_S$) and the breaking of 3-3-1 group to the SM is produced by the following hierarchical pattern:
 $${SU_L(3)\otimes U_X(1)}\stackrel{<v_\chi>}{\longrightarrow}{SU_L(2)\otimes
U_Y(1)}\stackrel{<v_\rho,v_\eta, v_S>}{\longrightarrow}{ U_{e.m}(1).}$$
The consistency of the model with the SM phenomenology is imposed by fixing a large scale for  $v_\chi$, responsible to give mass to the exotic particles  ($v_\chi \gg v_\rho, v_\eta, v_S$), with $v_\rho^2 + v_\eta^2 + v_S^2= v_W^2= \left( 246 \right)^2$ GeV$^2$.  

In the minimal version, the relation between $Z^\prime$, $V$ and $Y$ masses \cite{DION,NGL} is:
\begin{equation}
\frac{M_{V}}{M_{{Z^{\prime}}}} \simeq \frac {M_{Y}}{ M_{{Z^{\prime}}}} \simeq \frac{\sqrt{3-12\sin^2\theta_W}}{{2\cos\theta_W}}. 
\end{equation}
This special constraint respects the experimental bounds that, even being a consequence of the model,  is not often used in the literature. We keep this relation through our calculations.
For example, this ratio is $\simeq 0.3 $ for $\sin^2\theta_W =0.23$ \cite{PDG},  so that $Z^{\prime}$ can decays into a bilepton pair.

The interactions of quarks and neutral gauge bosons are described by the Lagrangian:
\begin{eqnarray}
&&{{\cal L}_{NC}}= \sum_{i} e q_i  \bar
\Psi_i\, \gamma^\mu  \Psi_i A_\mu - \frac{g}{2\cos\theta_W} \bigl\lbrace \bar
\Psi_i\, \gamma^\mu\ (g_{V_i} - g_{A_i}\gamma^5)\ \Psi_i\, Z_\mu 
\nonumber \\ &&
+ \bar\Psi_i\, \gamma^\mu\ (g^\prime_{V_i} - g^\prime_{A_i}\gamma^5)\ \Psi_i\, { Z_\mu^\prime}
\bigr\rbrace, 
\end{eqnarray}
\noindent where  $e q_i$ is the quark electric charge and 
$ g_{V_i}$, $g_{A_i}$, $g^\prime_{V_i}$ and $ g^\prime_{A_i}$ are the quark vector and axial-vector couplings with $Z$ and $Z^{\prime}$ respectively. 

As referred before, in the 3-3-1 model, one family must  transform  with respect to $SU(3)$  rotations  differently to the other two.  This requirement  manifests itself when we collect the quark currents  in a part with  universal coupling with  $Z^\prime$ similar to the SM and another part corresponding to the  non-diagonal  $Z^\prime$ couplings.
The transformation of these non diagonal terms, in the mass eigenstates basis,  leads to  the flavor changing neutral Lagrangian
\begin{eqnarray}
{\cal L}_{FCNC}= \frac{g\, \cos \theta_W }{\sqrt{3 -12 \sin^2 \theta_W }}\left(\bar U_L\,\gamma^\mu \, {\cal U}^{\dagger}_L \, B \, {\cal U}_L \, U_L +  \bar D_L \, \gamma^\mu \, {\cal V}^{\dagger}_L \, B  {\cal V}_L \, D_L \right) Z^{\prime}_\mu.
\end{eqnarray}
\noindent  where
$$ U_L =  \left( u \,\,\,\, c \,\,\,\,  t
\right)^T_L, \quad  D_L =  \left( d \,\,\,\,  s \,\,\,\, b \right)^T_L \quad {\hbox{and}} \quad  B = {\hbox{diag}}\ \left( 1 \ 0 \ 0\right).$$
The mixing matrices  ${\cal U}$ (for {\it up}-type  quark)  and  ${\cal V}$ (for {\it down}-type quark),  that give raise to the quark masses, come from the Yukawa Lagrangian and are related to the Cabibbo-Kobayashi-Maskawa matrix, as
\begin{equation}
{\cal U}^{\,u\,\dagger} {\cal V}^{\, d} = V_{CKM},
\end{equation}

By convention, in  the SM, it is usual   to assume that for {\it up}-type  quark the gauge interaction eigenstates are the same as the mass eigenstates, which corresponds to ${\cal U}^u=I$. This assumption is not valid in the 3-3-1 model because, in accordance to the renormalization group equations (RGE), all  matrix elements evolve with  energy and are unstable against radiative corrections. It turns out that ${\cal U}^u$ must be $ \not =  I$.  As the Eq. (9) is independent of representation, one is free to choose which quark family representation  must be different from the other two. We recall that our choice was for the first family to  belong to the triplet $SU(3)$ representation.  In the next section we will discuss the consequences of our choice.

All universal neutral couplings  diagonal and non-diagonal  are presented  in the Tables 1 and 2 respectively.
\begin{table}[ht]\label{sagui}
\begin{footnotesize}
\begin{center}
\begin{tabular}{||c|c|c||}
     \hline
\hline
&    &          \\ 
&  Vector Couplings & Axial-Vector Couplings    \\
&    &          \\ \hline
\hline
&    &      \\
$Z \bar u_i u_i$ & $\displaystyle{\frac{1}{2}-\frac{4\sin^2\theta_W}{3}}$ &
$\displaystyle{\frac{1}{2}}$  \\
&      &   \\ \hline
\hline
&       &   \\
$Z \bar d_j d_j$  &   $\displaystyle{-\frac{1}{2}+\frac{2\sin^2\theta_W}{3}}$  &
$\displaystyle{-\frac{1}{2}}$  \\
&     &     \\  \hline
\hline
&     &     \\
$Z^{\prime} \bar u_i u_i$ & $\displaystyle{\frac{1-4\sin^2\theta_W-{\cal U}^*_{ii} {\cal U}_{ii} \cos^2\theta_W}{\sqrt{3-12\sin^2\theta_W}}}$
& $\displaystyle{\frac{-1-4\sin^2\theta_W+{\cal U}^*_{ii} {\cal U}_{ii}\cos^2\theta_W}{\sqrt{3-12\sin^2\theta_W}}}$  \\
&    &    \\ \hline
\hline
&     &        \\
$Z^{\prime} \bar d_j d_j$ &  
$\displaystyle{\frac{1+2\sin^2\theta_W-{\cal V}^*_{jj} {\cal V}_{jj} \cos^2\theta_W}{\sqrt{3-12\sin^2\theta_W}}}$  &  $\displaystyle{\frac{-1+2\sin^2\theta_W+{\cal V}^*_{jj} {\cal V}_{jj} \cos^2\theta_W}{\sqrt{3-12\sin^2\theta_W}}}$  \\
&      &      \\ \hline
\hline
\end{tabular}
\end{center}
\end{footnotesize}
\caption{The $Z$ and $Z^{\prime}$ vector and axial-vector couplings to quarks ($u_1= u, u_2= c, u_3= t$, and $d_1= d, d_2= s, d_3= b$) in the Minimal Model; $\theta_W$ is the Weinberg angle and ${\cal U}_{ii}$ and ${\cal V}_{jj}$ are $\cal U$ and $\cal V$ diagonal mixing  matrix elements.}
\end{table}

\begin{table}[ht]\label{perro}
\begin{footnotesize}
\begin{center}
\begin{tabular}{||c|c|c||}
     \hline
\hline
&    &          \\ 
&  Vector Couplings & Axial-Vector Couplings    \\
&    &          \\ \hline
\hline
&    &      \\
$Z^{\prime} \bar c u$ & $\displaystyle{-\frac{{\cal U}^*_{12}\,{\cal U}_{11}\,cos^2\theta_W }{\sqrt{3-12\,\sin^2\theta_W}}}$ &
$\displaystyle{\frac{{\cal U}^*_{12}\,{\cal U}_{11}\,cos^2\theta_W }{\sqrt{3-12\,\sin^2\theta_W}}}$  \\
&      &   \\ \hline
\hline
&       &   \\
$Z^{\prime} \bar t u$  &   $\displaystyle{-\frac{{\cal U}^*_{13}\,{\cal U}_{11}\,cos^2\theta_W }{\sqrt{3-12\,\sin^2\theta_W}}}$  &
$\displaystyle{\frac{{\cal U}^*_{13}\,{\cal U}_{11}\,cos^2\theta_W }{\sqrt{3-12\,\sin^2\theta_W}}}$  \\
&     &     \\  \hline
\hline
&     &     \\
$Z^{\prime} \bar t c$ & $\displaystyle{-\frac{{\cal U}^*_{13}\,{\cal U}_{12}\,cos^2\theta_W }{\sqrt{3-12\,\sin^2\theta_W}}}$
& $\displaystyle{\frac{{\cal U}^*_{13}\,{\cal U}_{12}\,cos^2\theta_W }{\sqrt{3-12\,\sin^2\theta_W}}}$  \\
&    &    \\ \hline
\hline
&     &        \\
$Z^{\prime} \bar d s$ &  
$\displaystyle{-\frac{{\cal V}^*_{12}\,{\cal V}_{11}\,cos^2\theta_W }{\sqrt{3-12\,\sin^2\theta_W}}}$  &  $\displaystyle{\frac{{\cal V}^*_{12}\,{\cal V}_{11}\,cos^2\theta_W }{\sqrt{3-12\,\sin^2\theta_W}}}$  \\
&      &      \\ \hline
\hline
&       &   \\
$Z^{\prime} \bar b d$  &   $\displaystyle{-\frac{{\cal V}^*_{13}\,{\cal V}_{11}\,cos^2\theta_W }{\sqrt{3-12\,\sin^2\theta_W}}}$  &
$\displaystyle{\frac{{\cal V}^*_{13}\,{\cal V}_{11}\,cos^2\theta_W }{\sqrt{3-12\,\sin^2\theta_W}}}$  \\
&     &     \\  \hline
\hline
&       &   \\
$Z^{\prime} \bar b s$  &   $\displaystyle{-\frac{{\cal V}^*_{13}\,{\cal V}_{12}\,cos^2\theta_W }{\sqrt{3-12\,\sin^2\theta_W}}}$  &
$\displaystyle{\frac{{\cal V}^*_{13}\,{\cal V}_{12}\,cos^2\theta_W }{\sqrt{3-12\,\sin^2\theta_W}}}$  \\
&     &     \\  \hline
\end{tabular}
\end{center}
\end{footnotesize}
\caption{The flavor changing vector and axial-vector couplings to quarks ($u$- and $d$-type ) induced by $Z^{\prime}$ in the Minimal Model.}
\end{table}

The dominant  couplings between  ordinary to exotic quarks are driven by single charged bilepton as follows:  
\begin{eqnarray}
{\cal L}_{CC}= - \frac{g}{2 \sqrt 2} \bigl \lbrack \overline d \gamma^\mu(1 - \gamma^5)\left({\cal V}_{21} \,\,j_2 + {\cal V}_{31} \,\,j_3\right) + \overline J_1 \gamma^\mu(1 - \gamma^5)\,\, {\cal U}_{11} \,\,u \bigr \rbrack V^+_\mu.
\end{eqnarray}
\noindent where ${\cal V}_{21}$, ${\cal V}_{31}$ and ${\cal U}_{11}$ are mixing matrices elements (Eq. (9)).

\par
In addition to the SM gauge boson Lagrangian the trilinear terms used in the present work are: 
\begin{eqnarray}
&&{\cal L}_{gauge}=-i g \sin \theta_W \left[ A^{\nu} (
V^-_{\mu \nu} V^{+\mu} - V^+_{\mu \nu} V^{-\mu} ) +
A_{\mu \nu} V^{-\mu} V^{+\nu}\right] \nonumber \\ 
 & &  + \frac{i g}{2} \left(\cos\theta_W + 3 \sin\theta_W \tan\theta_W \right) \left[ Z^\nu (
V^-_{\mu \nu} V^{+\mu} - V^+_{\mu \nu} V^{-\mu}  +
Z_{\mu \nu} V^{-\mu} V^{+\nu}\right]  \nonumber \\ 
& & + \frac {i g} {2} \sqrt{3(1 - 3 \tan^2\theta_W) } \left[ Z^{\prime \nu }(
V^-_{\mu \nu} V^{+\mu} - V^+_{\mu \nu} V^{-\mu} ) +
Z^\prime_{\mu \nu} V^{-\mu} V^{+\nu}\right],
\end{eqnarray}
where $B_{\mu  \nu} \equiv  \partial_\mu B_\nu -
\partial_\nu  B_\mu$, with $B = A, Z, Z^\prime$ and $ V^\pm$.
\par

Finally, one of the main features of the model comes from the relation between the  $SU_L(3)$ and $U_X(1)$ couplings, expressed as:
\begin{equation}
\frac {g^{\prime\, 2}}{g^2} =\frac{\sin^2 \theta_W}{1\, -\, 4 \sin^2 \theta_W}.
\end{equation}
that fixes $\sin^2 \theta_W < 1/4$, which is a peculiar characteristic of this model, as 
explained in the Introduction section.

\section{Results}

In this paper we focus on the bilepton ($V^{\pm}$) pair production in $pp$ collision at LHC. This particle is predicted in many extensions of the SM and in particular in the 3-3-1 model that was used in the present paper. We restrict our calculation to a version of the model where the bilepton mass is related to the mass of the extra neutral gauge boson $Z^\prime$ also predicted in the model, by the Eq. (6).

 We fix the exotic quark masses ($J_1, j_2$ and $j_3$)  to be $600$ GeV and for the $V^{\pm}$ mass we use  a set of values compatible with the findings related to the $Z \longrightarrow b \bar b $ \cite{GON}, where  the authors obtained the allowed region for  exotic quark and bilepton masses, through the deviation between the SM calculation and the experimental data.  
We adopt the $Z^\prime $ mass in the range from $800$ to $1200$ GeV, which is in accordance with  accepted bounds \cite{PDG}.  All these values are shown in the Table 3. 

 The group structure of model is such that bileptons couple ordinary to exotic quarks and  leptons ($e$, $\mu$ and $\tau$) with their neutrinos. In the hadronic channel the bilepton can decay in $d-$type quark with  $j_{2,3}$ and $u-$type quark with $J_1$. However, for the range of extra neutral  gauge boson mass considered here the only decay mode is leptonic  because
the exotic-quark ordinary-quark channel will only opens when $M_{Z^{\prime}}= 2 $ TeV, associated with a bilepton heavier than $600$ GeV.
   In contrast with $W$, which decays into  $\bar \nu_\ell + \ell$ where $ \ell$ is emitted softly, the leptons coming from bilepton carry  high transverse momentum. This signature can be used to disentangle the processes of bilepton pair production from  $W$ pair production. 

In order to calculate the total cross section for bilepton pair production we start by  considering the  elementary process, $ q_i + \bar q_i \lra V^+ + V^- $ ($q_i = u, d$), taking into account all contributions: $\gamma$, $Z$ and a new neutral gauge boson ${Z^\prime}$ in the  $s-$ channel and exotic heavy quarks  $Q_j$ ($J_1, j_2$ and $j_3$) in the  $t-$ channel, as displayed in the Figure 1.  At the beginning of our calculation, we have taken into account the heavy quark  and $Z^{\prime}$  widths, however we have verified that our results do not depend on the heavy quarks width then we keep only $Z^{\prime}$ width in all calculations. We perform the amplitude algebraic calculation with FORM \cite{VER}.

The elementary differential cross section obtained, as a function of kinematical invariants ($\hat s$, $\hat t$, $\hat u$), can be computed, for 
 $k, l = \gamma$, $Z$, $Z^\prime$ and $Q_j = J_1, j_2$ and  $j_3$, as: 

\begin{equation}
\frac {d\hat \sigma} { d \hat t}= \frac {2\pi\alpha^2} { \hat s^2} 
\sum_{k l}{B_{k l}}
\end{equation}
We present below  the amplitudes ($B_{k l}$) corresponding to the  diagrams shown in the Figure 1:
\begin{eqnarray}
&&B_{\gamma \gamma}=(eq_i)^2 A(\hat s,\hat t, \hat u)\nonumber\\
&&B_{Z Z}=\Bigl(\frac{e_Z}{e^2}\Bigr)^2\left[g_V^2+g_A^2\right]
\, \Delta_Z^2\,  A(\hat s,\hat t,\hat u)\nonumber\\
&&B_{\gamma Z} =  -2 eq_i\left(\frac{e_{Z}}{e^2}\right)
g_V\,  \Delta_Z\,  A(\hat s,\hat t,\hat u)\nonumber\\
&&B_{Z^{\prime}Z^{\prime}} =\Bigl(\frac{e^{\prime}_Z}{e^2}\Bigr)^2 \left(g_V^{\prime 2}+g_A^{\prime 2}\right)\, \Delta_{Z^\prime}^2\,  A(\hat s,\hat t,\hat u)\nonumber\\
&&B_{Z^{\prime} Q_j}  =  2sgn(eq_i) \frac{G_{VA}^{\prime 2}}{e^2}
\frac{e_Z}{e^2}\left[g_V^{\prime}\left(a^2_j+b^2_j\right)
 -2a_jb_j g_A^{\prime}\right]\, \Delta_{Z^\prime}\, 
I_{1}(\hat s,\hat t,\hat u)\nonumber\\
&&B_{Z Z^{\prime}} =  - \Bigl(\frac{e_Z}{e^2}\Bigr)\Bigl(\frac{e^{\prime}_Z}{e^2}\Bigr)(g_A g^{\prime}_A+ g_Vg^{\prime}_V) \, \Delta_Z \Delta_{Z^\prime}\, A(\hat s,\hat t,\hat u)\nonumber\\
&&B_{\gamma Q_j} =  -2 eq_i sgn(eq_i)\frac{G_{VA}^{\prime 2}} {e^2}
\left[a^{2}_j+b^{2}_j\right]I_{1}(\hat s,\hat t,\hat u)\nonumber\\
&&B_{Q_j Q_j} =  \frac{G_{VA}^{\prime 4}}{e^4}
\left[\left(a^4_j+b^4_j+6\left(a_j b_j\right)^2\right)
E_{2}(\hat s,\hat t,\hat u)+ M_{Q_j}^2\left(a_j^2-b_j^2\right)^{2}
E_{H}(\hat s,\hat t,\hat u)\right]\nonumber\\
&&B_{Z Q_j}  =  2sgn(eq_i) \frac{G_{VA}^{\prime 2}}{e^2}
\frac{e_Z}{e^2}\left[g_V\left(a^2_j+b^2_j\right)
 -2a_jb_j g_A\right]\, \Delta_Z\, 
I_{1}(\hat s,\hat t,\hat u)\nonumber\\
&&B_{\gamma Z^{\prime}} =  -2 eq_i \left(\frac{e^{\prime}_{Z}}{e^2}\right) g_V^{\prime} \, \Delta_{Z^\prime}\,  A(\hat s,\hat t,\hat u)
\end{eqnarray}

\nl where $M_Z$, $M_Z^\prime$ are the neutral gauge boson masses and $\Gamma_{Z^\prime}$ is the ${Z^\prime}$ width;  $a_j $ and $b_j$ are the  ordinary quark-exotic quark-bilepton couplings ($a_j = b_j =1$), $G_{VA}^{\prime} = \displaystyle \frac {g}{2 \sqrt 2} {\cal V}^{u, d}_{i j}$, $\displaystyle  {\Delta_Z=\frac{\hat s}{{\hat s}-M^2_Z}}$ and 
$\displaystyle {\Delta_{Z^\prime}=\frac{\hat s}{\hat s - M^2_{Z^\prime}+ M_{Z^\prime}\Gamma_{Z^\prime}}}$.
\sk

\nl   The  trilinear coupling constants $e_Z$ and $e_Z{^\prime}$ are:
$$e_Z =  \displaystyle{ \frac{e}{2}\frac{1+2\sin^2\theta_W}{\sin\theta_W\cos\theta_W}} \quad {\hbox {and}} \quad  e_{Z^{\prime}} = \displaystyle{ \frac{e}{2}\frac{\sqrt{3-12\sin^2\theta_W}}{\sin\theta_W\cos\theta_W}}.$$
\bigskip
The functions $A(\hat s,\hat t,\hat u)$,
$E(\hat s,\hat t, \hat u)$ and $I(\hat s, \hat t, \hat u)$ are:
\begin{eqnarray}
&&A(\hat s,\hat t,\hat u)= \left(\frac{\hat u\, \hat t}{M^4_V} -1\right)
\left(\frac{1}{4}- \frac {M^2_V}{\hat s} + \frac {3\,M^4_V}{\hat s^2}\right)+
\frac {\hat s} { M^2_V} - 4 \nonumber\\
&&I(\hat s,\hat t,\hat u)=\left(\frac{\hat u \,\hat t }{M^4_V}-1\right)
\left(\frac 1 4 - \frac {M^2_V} {2\,\hat s}
- \frac {M^4_V} {\hat s\, \hat t}\right) +
\frac {\hat s}{M^2_V}  +  \frac {2\,M^2_V}{\hat t} - 2\nonumber\\
&&E(\hat s,\hat t,\hat u)=\left(\frac {\hat u \,\hat t}{ M^4_V}-1\right)
\left(\frac 1 4  + \frac {M^4_V}{\hat t^2} \right) + \frac {\hat s}{ M^2_V},
\end{eqnarray}

\nl and the functions originated by exotic quark contributions are:
\begin{eqnarray}
&&E_1(\hat s,\hat t,\hat u)={{\hat t}\over{{\hat t}-M^2_{Q}}}
E(\hat s,\hat t,\hat u)\nonumber\\
&&I_1(\hat s,\hat t,\hat u)={{\hat t}\over{{\hat t}-M^2_{Q}}}
I(\hat s,\hat t,\hat u)\nonumber\\
&&E_2(\hat s,\hat t,\hat u)=\biggl({{\hat t}\over{{\hat t}-M^2_{Q}}}
\biggr)^2 E(\hat s,\hat t,\hat u)\nonumber\\
&&E_H(\hat s,\hat t,\hat u)=\biggl({{\hat t}\over{{\hat t}-M^2_{Q}}}
\biggr)^2 \left[{M^2_V \over {\hat t^2}}\left({{\hat u\, \hat t}
\over {M^4_V}}-1\right)+{{\hat s}\over{M^4_V}}\left({1\over 4}+
{M^4_V \over {\hat t^2}}\right)\right].
\end{eqnarray}
\par

We have performed the $\hat t$ integration within the limits:  $ \hat t_{min}=M_V^2+M_q^2-\hat s/2(1-\beta(-1+\cos\theta))$ and 
$\hat t_{max}=M_V^2+M_q^2-\hat s/2(1-\beta(1-\cos\theta))$, where  $M_V$, $M_q$ and $M_{Q}$ are the bilepton, ordinary and exotic quark masses and the definition 
$$\beta= \frac{\sqrt{(\hat s-4\,M_V^2)(\hat s-4\,M_q^2)}}{\hat s}.$$

\par
For short we call ${\cal X} = \int X(\hat s,\hat t,\hat u) \,\,d\hat t $, where ${ X}= {A}, \, {E}, \,{ I}, \, { I}_1,\,  { E}_1, \,  {E}_2, \,  { E}_H $ and ${\cal X}= {\cal A}, \, {\cal E}, \,{\cal I}, \, {\cal I}_1,\,  {\cal E}_1, \,  {\cal E}_2, \,  {\cal E}_H $:
\begin{eqnarray}
&&{\cal A} =
\beta\, \left( s-4\,{M^2_V} \right)  \left( \frac{1}{2}+ {\frac {5\,s}{
{6\,M^2_V}}}+ {\frac {s^2}{24\,{M^4_V}}} \right) 
\nonumber\\
&&{\cal E} = \beta\,s \left( - 2 
 + {\frac {5\,s}{{6\,M^2_V}}} +  {\frac {s^2}{{24\,M^4_V}}} \right) + {\left( s -2\,{M^2_V}\right)}\,  L_{\chi} \nonumber\\
&&{\cal I}=\beta\, \left( s-2\,{M^2_V}\right)  \left(\frac{1}{2}+ {\frac {5\,s}{{6\,M^2_V}}}+ {\frac {s^2}{24\,M^4_V}} \right)
-{\frac { \left( 2\,s +{M^2_V} \right){M^2_V} }{s}}\, L_{\chi}
 \nonumber\\
\end{eqnarray}
\noindent and

\begin{eqnarray*}
&&{\cal E}_1=-\frac{\beta\, s}{4M^4_V}\left[{{M^4_{Q}}} +{\left(\frac{s}{2} - M^2_V  \right) {M^2_{Q}}} + \left( 4 {M^4_V} -3 s M^2_V+ \frac{{s^2}}{12}({\beta}^2 - 3) \right) \right]
  -\frac {{M^4_V} }{M^2_{Q}} L_{\chi}\cr
&&+ \frac{1}{4 {M^4_V}}\left[M^6_{Q}+( s-2 {M^2_V}) M^4_{Q}+ \left ( 5 {M^4_V}-4 {M^2_V} s \right) M^{2}_
{Q}- 4 (2 M^2_V - s){M^4_V}+\frac {M^8_V}{M^2_{Q}}
\right ] L_{\psi} 
\eqn
\beqn 
&&{\cal E}_2=\left[ {\frac {{M^6_Q}}{M^4_V}}+ \left( -\frac{3}{2\,M^2_V}+\,{\frac {3\,s}{4\,M^4_V}} \right) {M^4_{Q}}
+ \left( -{\frac {2\,s}{{M^2_V}}}+\frac{5}{2} \right) {M^2_Q}
+s-2\,{M^2_V} \right] {L_{\psi}}\cr
&&
-\frac{\Delta}{3}
 \left[ 24\,M^8_Q + 30\,\left(\,s -2\, M^2_V
 \right) M^6_Q + \left( 96\,M^4_V -68\,M^2_V
\,s+5\,s^2 \right) M^4_Q \right. \cr && \cr && \left. 
-\left( s^3 - 94\,M^4_V \,s+ 18\,M^2_V\,s^2+108\,M^6_V \right) M^2_Q  -M^4_V\,s^2+48\,M^8_V -20\,M^6_V\, s \right] 
\eqn
\beqn
&&{\cal E}_H= \frac{\Delta}{2} \left[ 8\left( s-4\,M^2_V \right) 
M^4_Q + 4\left( 16\,M^4_V-10\,M^2_V\,s+\,s^2
 \right) M^2_Q - 32 \, M^6_V - 8\, M^2_V \, s^2 \right. \cr && \cr && \left. 
+ 36 \, M^4_V \, s + s^3 - s^3 \, \beta^2 + 4\, M^2_V \, s^2\, \beta^2 \right]  - \frac{1} {2\,M^4_V}  \,\left( 4\,M^4_V - 4\,M^2_Q\,M^2_V+ M^2_Q\,s -
2\,M^2_V \, s \right){L_{\psi}}
\eqn
\begin{eqnarray}
&&{\cal I}_1=
{\frac { \left(s - 2\,{M^2_V} \right){M^6_{Q}}}{4\,M^4_V\,s}}{L_{\psi}}\,-\left[\beta\,(s-2\,{M^2_V})-(s-4\,M^2_V)L_{\psi} \right] \frac{M^4_{Q}}{{4\,M^4_V}}\cr
&& + \left[ 2\,\beta\,\left(M^4_V-\frac{s^2}{4}-M^2_V\,s\right)-\frac { M^2_V
\left(4\,{s}^{2}-6\,{M^4_V}-5\,{M^2_V}\,s \right)}{s}\, L_{\psi} \right]\frac{M^2_{Q}}{4 M^4_V}\cr
 && + \frac{\beta}{4 M^4_V} \left(\frac{s^{3}}{6}+3 M^2_V  s^{2}-\frac{14 M^4_V s}{3}-
4 M^6_V\right) -{\frac {48\, {M^2_V}
\left( 2 s+{M^2_V} \right) }{s}} {L_{\psi}}
\end{eqnarray}

 \noindent where: $$\displaystyle{L_{\chi}= 2\,\ln  \left( {\frac {1+ \beta}{1- \beta}} \right)},  \qquad  \displaystyle{L_{\psi} = \ln \left(   
{\frac {2\,{M^2_Q}-2\,{M^2_V}+s+s \beta}
       {2\,{M^2_Q}-2\,{M^2_V}+s-s\beta}   } \right) }$$
and 
$$ \displaystyle{\Delta=\frac{s\,\beta}{8\,M^4_V \left( M^4_Q + \left( s-2\,M^2_V
 \right) M^2_Q + M^4_V  \right) }}.$$
The crossing relation for the  functions coming from $\hat t$ and $\hat u$ channels gives,
\begin{eqnarray*}
\int  E(\hat s,\hat t,\hat u) \, d\hat t &=&\int  E(\hat s,\hat u,\hat t)\, d\hat u\nonumber\\
\int I(\hat s,\hat t,\hat u) \,  d\hat t &=&\int I(\hat s,\hat u,\hat t) \,d\hat u\nonumber\\
\end{eqnarray*}
In order to obtain the elementary total cross section for each  parton in the initial proton, we  sum the contributions ${B_{lm}}$ integrated over $\hat t$,
$$ \hat\sigma_{q\bar q} = \int \frac{d \hat\sigma}{d \hat t} d \hat t =\frac {2\pi\alpha^2}{\hat s^2} \sum_{lm} \int\, {B_{lm} \, d \hat t}$$ 

It is well known that in the SM  the correct behavior of the total cross  section at high energy for charged gauge boson pair production is extremely dependent on the balance between {\it s-} and {\it t-}channel contributions \cite{JOH}, as can be seen in the Figure 2, which shows the elementary cross sections ($u \bar u$ and $d \bar d$) for $W$ pair production. In fact, the renormalizability of the theory is translated into the cancellation between these contributions at high energy. For some models the new neutral boson and/or exotic fermion contributions can guarantee this delicate cancellation \cite{VDM}. 

Let us discuss how this comes about for the present  model where there are  two non standard contributions: $Z^\prime$ and exotic quarks. We present in Figures 3 and 4 some individual amplitudes $B_{ij}$ (Eq. 14) that correspond  to the elementary  sub-processes, $u \bar u$ and $d \bar d$ for $M_Z{\prime}=800$ GeV.  In the Figure 3 we plot the main contributions ($B_{Z^{\prime}Z^{\prime}}$,  $B_{Z Z}$, $B_{\gamma \gamma}$, $B_{Z Z^{\prime}}$,  $B_{\gamma Z}$,  $B_{Q_j Q_j}$). We do not display the remaining interferences ($B_{\gamma Z^{\prime}}$,  $B_{Z Q_j }$, $B_{\gamma Q_j}$, 
$B_{ Z^{\prime} Q_j}$), that get values between  $B_{\gamma Z^{\prime}}$ and   $B_{Q_j Q_j}$. In the Figure 4, we show the similar relevant contributions for 
$d \bar d$ with one  additional exotic quark. We  omit the curves corresponding to 
 $B_{\gamma Z^{\prime}} \simeq B_{\gamma \gamma}$ ,  $B_{Z Q_1}+ B_{Z Q_2}\simeq  B_{Q_1 Q_1 } + B_{Q_2 Q_2 }+ B_{Q_1 Q_2}$, $B_{\gamma Q_1} + B_{\gamma Q_2} \simeq B_{\gamma Z}$, $B_{Z^{\prime} Q_1} + B_{Z^{\prime} Q_2} <  B_{Z Z^{\prime}}$.  One can observe again the large extra gauge boson ($B_{Z^{\prime}Z^{\prime}}$) component and the tiny exotic quark and interference contributions.  

Let us show in the Figure 5 the bad behavior in energy for the elementary
cross section when we consider only the neutral boson contributions ($\gamma$, $Z$ and $Z^\prime$). One can see clearly that, when the energy increases, the $u \bar u$ sub-process violates
softly the unitarity behavior (only visible for $\sqrt {\hat s} > 4$ TeV) and, on the
other hand, $d \bar d$ leads to a more drastic violation. As expected, the $d \bar d$  process
is more sensitive than $u \bar u$, because $d \bar d$  channel receives more exotic quark
contribution. This behavior imposes to add the exotic quark contributions ({\it via} t) involving the charged current. The amplitudes balance has to occur between the exotic quark ({\it via}  t) and the $s-$channel. 

Besides, this cancellation requires to take into account the mixing between the quark eigenstates respecting the constraint given by Eq. (9). The quark mixing depends on  what family must belong to a different $SU(3)_L$ representation. Working in the minimal version \cite{PIV}, where the first family is in the $SU(3)$ triplet representation, it is possible to obtain mixing parameters   compatible with Eq. (9) and restoring the correct high energy behavior.
There is no parametrization, in the literature, for ${\cal U}$ matrix elements, however some limits  for ${\cal V}$ elements have been obtained  from $Z^{\prime}$ rare decay bounds in \cite {PRO,SHE,GOM}.
We cannot exclude the possibility that another choice of mixing parameters would provide a correct energy behavior, when the third quark family is treated differently. 

The Figures 6 and 7 show the elementary   $ q + \bar q\lra V^+ + V^-$ total cross section for the  set of quark mixing parameters: ${\cal U}_{11} = 0.1349989$, ${\cal V}_{11} = 0.900542$, ${\cal V}_{12} = 0.1009984$ and ${\cal V}_{31}={\cal V}_{11}$. In these figures 
 we have used three values for $M_{Z^\prime}= 800 $ GeV, $1000$ GeV and $1200$ GeV for $M_Q= 600 $ GeV. It can be noted the good behavior of the elementary cross sections presenting a peak  around the ${Z^\prime}$ mass and becoming broader and smaller as ${Z^\prime}$ mass increases.
This range of values relies on our  previous work \cite{EUR} where we have establish bounds on $Z^{\prime}$ mass in two versions of 3-3-1 models for $e^+ e^-$ and  hadron colliders, obtaining results that are compatible with experimental bounds. Here we consider  the constraints given by Eq. (6).  We  display  in the Table I some values for $M_{Z^\prime}$, $\Gamma_{Z^\prime}$, $M_V$ and $\Gamma_V$. 

The total cross section for $ p + p \lra V^+ + V^- + X$ is obtained integrating the elementary total cross section weighted by the distribution function for partons in hadron (proton) \cite{CTE}.
$$
\sigma(p + p  \lra V^+ + V^- + X)=  \sum_{i, j} \int_{thr} ^1 \int_{thr} ^1 dx_1 dx_2 f_i(x_1, Q^2) f_j(x_2, Q^2)\ 
\hat \sigma_{qq}.$$
To obtain more realistic results, we have applied an angular cut on the angle between the final bileptons with respect to the initial beam direction, $\vert \eta\vert \le 2.5$. 

The final results are displayed in the Figure 8, where the total cross section is plotted as a function of the bilepton mass.  We consider two energy regimes ($\sqrt s = 10$ TeV and $\sqrt s = 14$ TeV)
to compare with the $W^+\, W^-$ production. It is clear  from the plot that, even for  $\sqrt s = 10$ TeV, the production of bileptons pairs with  $M_V \, \le 300$ GeV is larger than $W$ pair production,  allowing for a large number of events originated from bilepton decay. For a low LHC luminosity around $1\, fb^{-1}$ and c.m. energy of $10$ TeV it can be produced a thousand of  $M_V \simeq 300$ GeV pairs. For $\sqrt s = 14$ TeV the same number of pair  can be produced for $M_V \, \simeq 450$ GeV.  This scenario is related to a very massive extra neutral gauge boson existence.

\begin{figure} 
\includegraphics[height=.5\textheight]{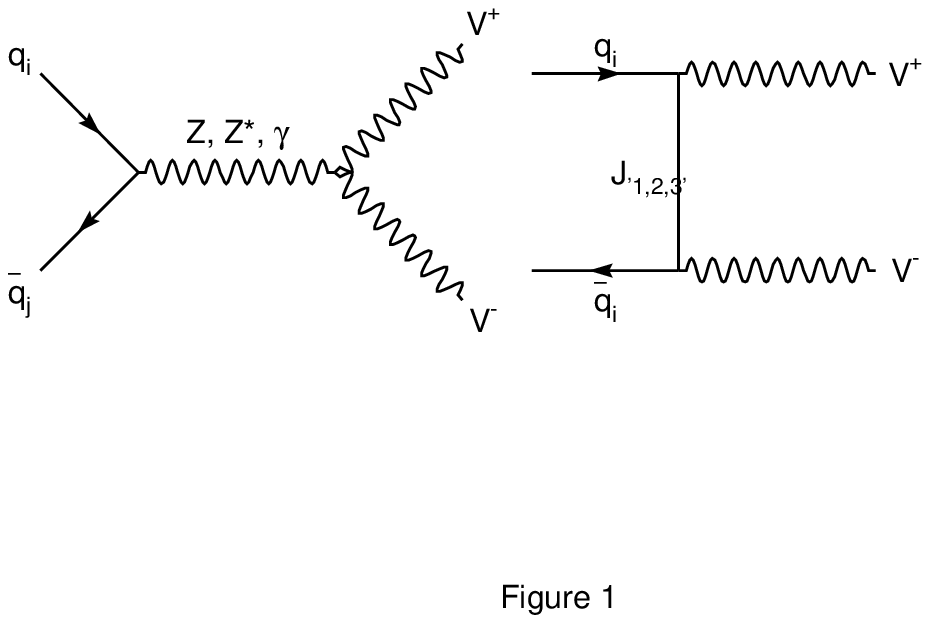}
\caption{The Feynman diagrams for $q + \bar q \lra V^+ + V ^-$ process with s-channel and t (u)-channel contributions.}
\end{figure}

\begin{figure} 
\includegraphics[height=.5\textheight]{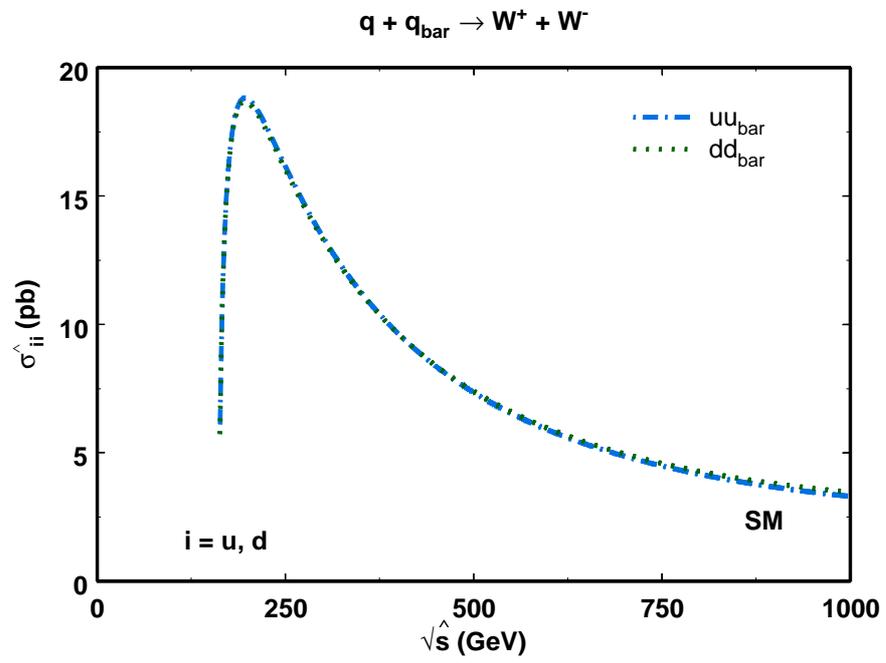}
\caption{The elementary total cross section for the process $ q + \bar q\lra W^+ + W^-$ for the SM.}
\end{figure}

\begin{figure} 
\includegraphics[height=.5\textheight]{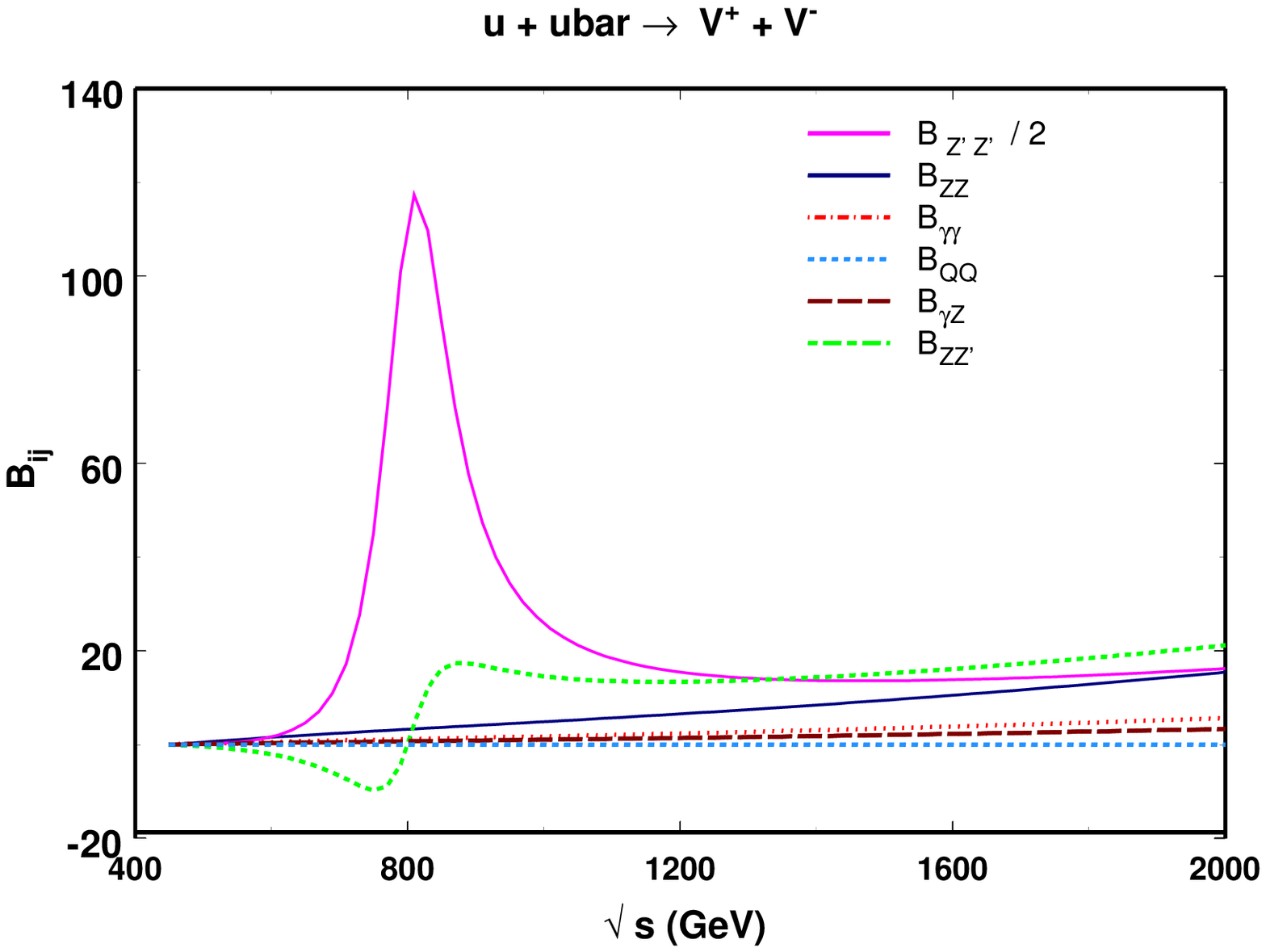}
\caption{The partial amplitudes the process $ u + \bar u\lra V^+ + V^-$ for the 3-3-1 model considering $M_{Z^\prime}= 800$ GeV and $M_{{J_1}}= 600 $ GeV.}
\end{figure}

\begin{figure} 
\includegraphics[height=.5\textheight]{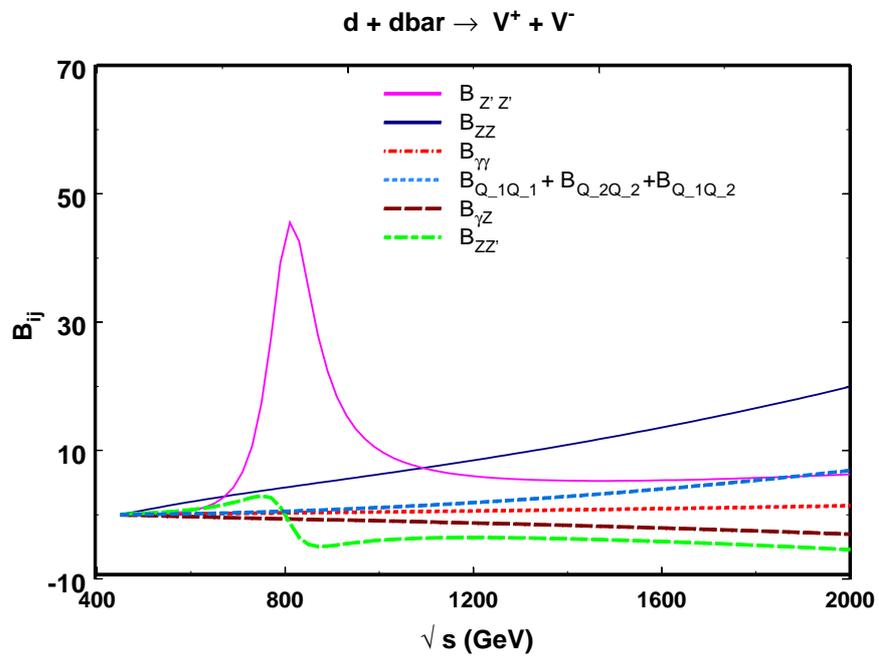}
\caption{The partial amplitudes the process $ d + \bar d\lra V^+ + V^-$ for the 3-3-1 model considering $M_{Z^\prime}= 800$ GeV and $M_{{j_2}}=M_{{j_3}}= 600 $ GeV.}
\end{figure}

\begin{figure} 
\includegraphics[height=.5\textheight]{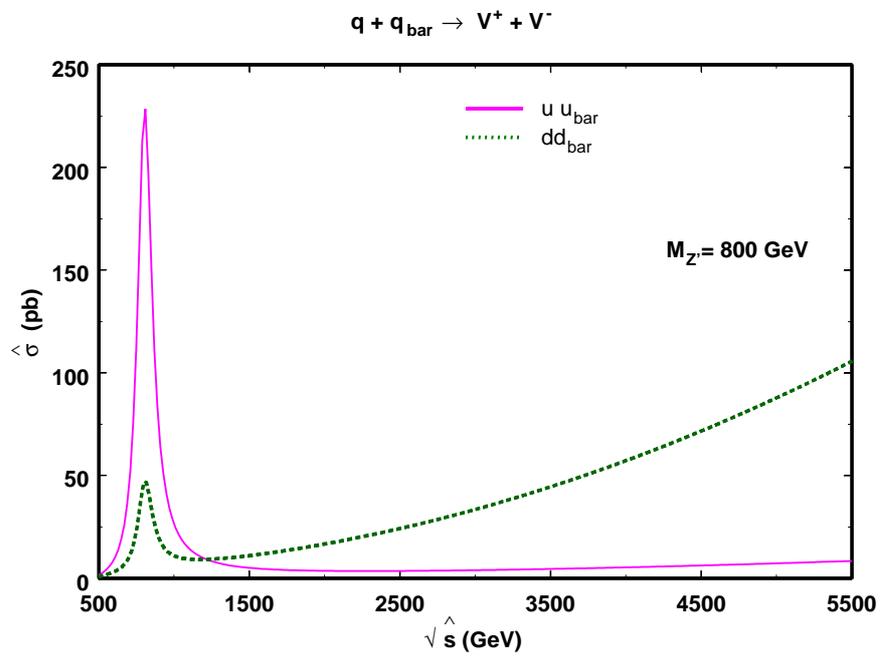}
\caption{The elementary cross section for $u \bar u$ and $d \bar d$ sub-process for $M_{Z^{\prime}}= 800 $ considering only the $s-$channel contributions.}
\end{figure}

\bigskip
\par

\begin{table}[ht]\label{manga}
\begin{center}
\begin{tabular}{||c|c|c|c||}
\hline \hline
 &  & &\\
$M_{Z^{\prime}}$ (GeV) &  $\Gamma_{Z^{\prime}}$ (GeV) &  $M_{V^\pm}$ (GeV) & $\Gamma_{V^{\pm}}$ (GeV)\\
 &  & & \\ \hline
\hline
&  &  & \\
$800$  &  $117$ & $217$ & $1.84$\\
&   &  & \\ \hline
\hline
&  &  & \\
$1000$ & $149$ & $271$ & $2.28$ \\
&  & & \\ \hline
\hline
&  &  &\\
$1200$ & $181$ & $407$ & $3.44$\\
&  & & \\ \hline
\hline
\end{tabular}
\end{center}
\caption{Some widths for new gauge bosons $Z^{\prime}$ and $V^{\pm}$ in the minimal 3-3-1 model.}
\end{table}

\begin{figure} 
\includegraphics[height=.5\textheight]{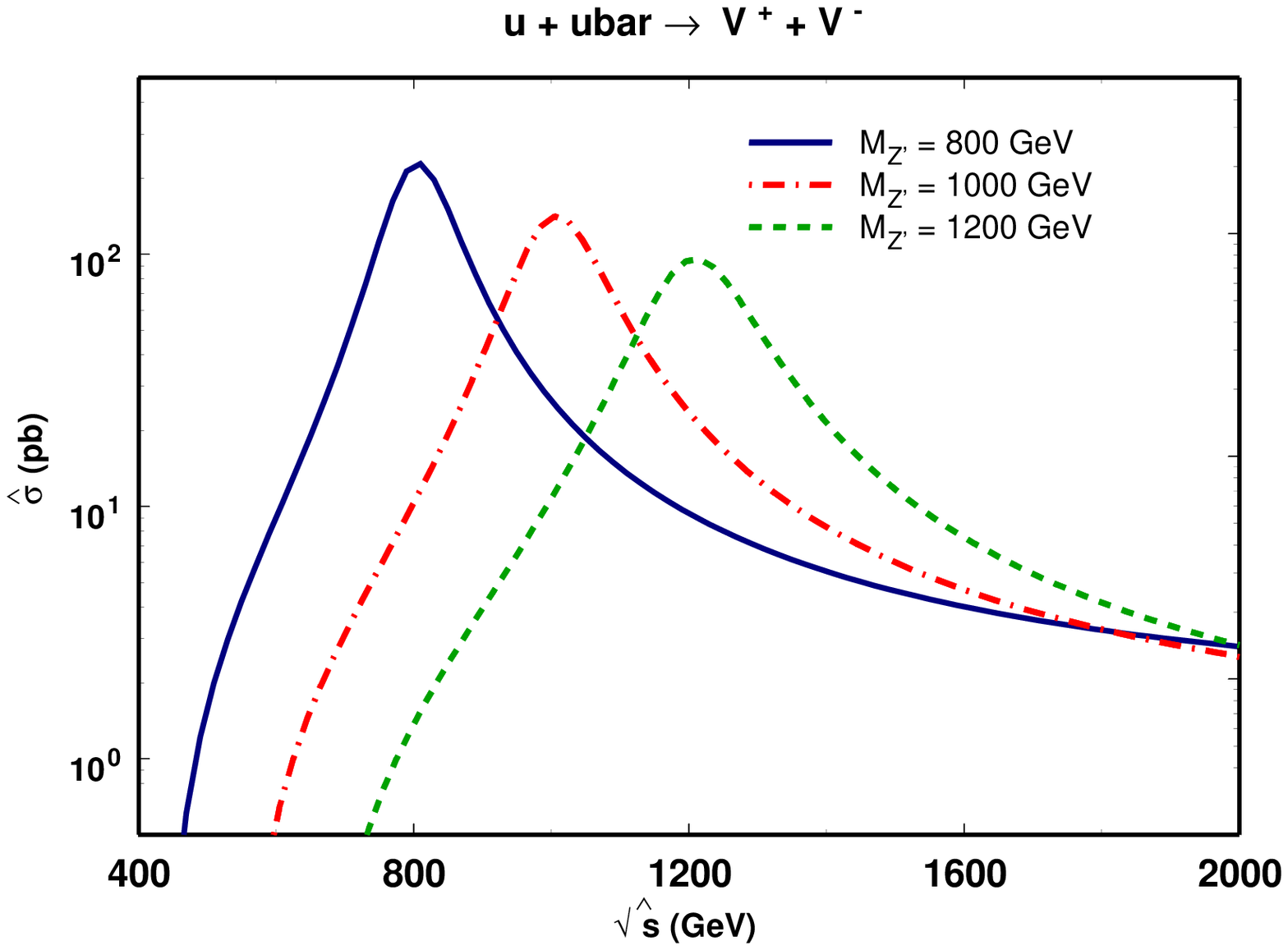}
\caption{The elementary total cross section for the process $ u + \bar u\lra V^+ + V^-$ for the 3-3-1 model considering $M_{Z^\prime}= 800 $ GeV, $1000$ GeV and $1200$ GeV and $M_{J_1}= 600 $ GeV.}
\end{figure}

\begin{figure} 
\includegraphics[height=.5\textheight]{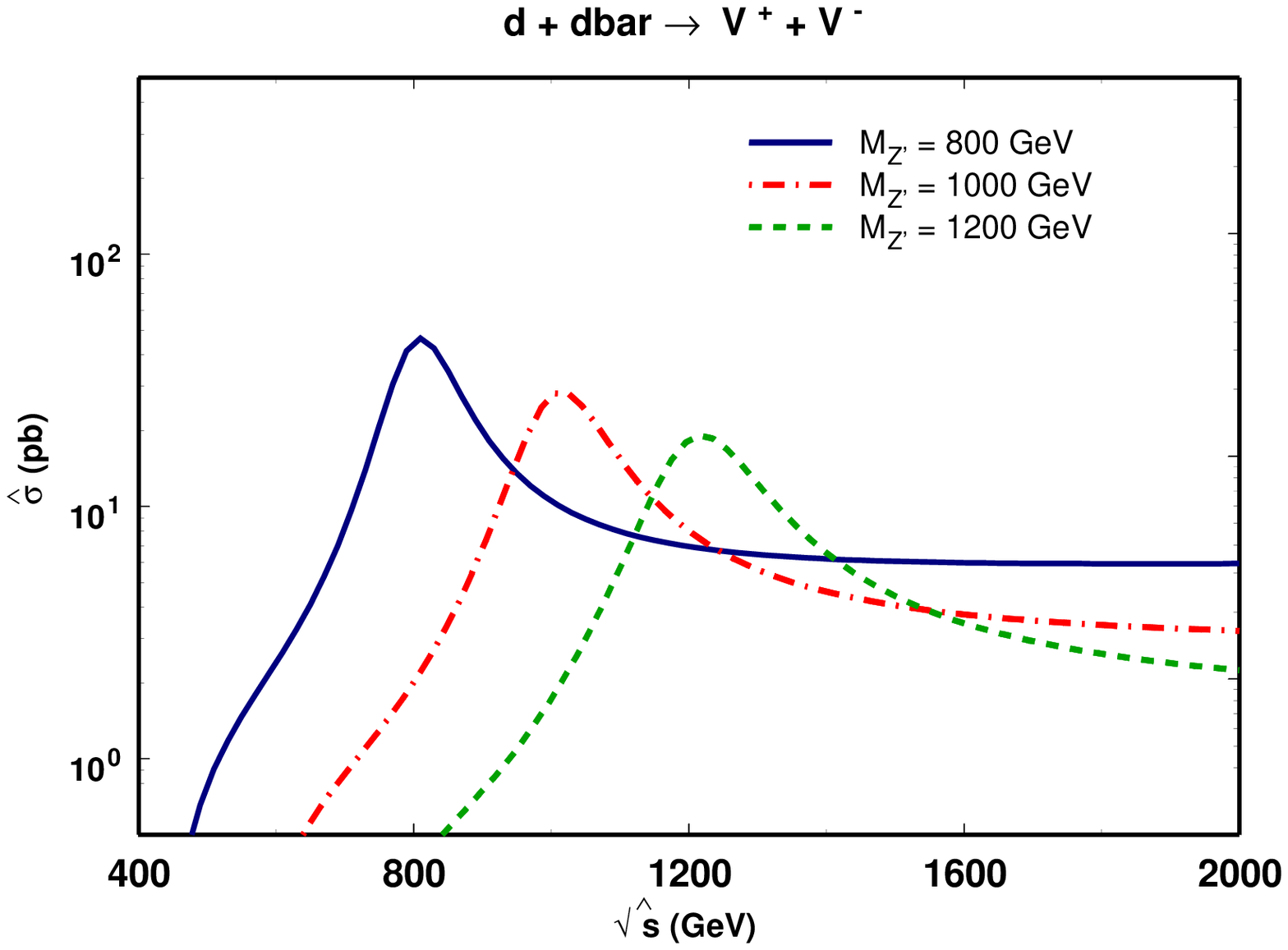}
\caption{The elementary total cross section for the process $ d + \bar d \lra V^+ + V^-$ for the 3-3-1 model considering $M_{Z^\prime}= 800 $ GeV, $1000$ GeV and $1200$ GeV and $M_{{j_2}}=M_{{j_3}}= 600 $ GeV.}
\end{figure}

\begin{figure} 
\includegraphics[height=.5\textheight]{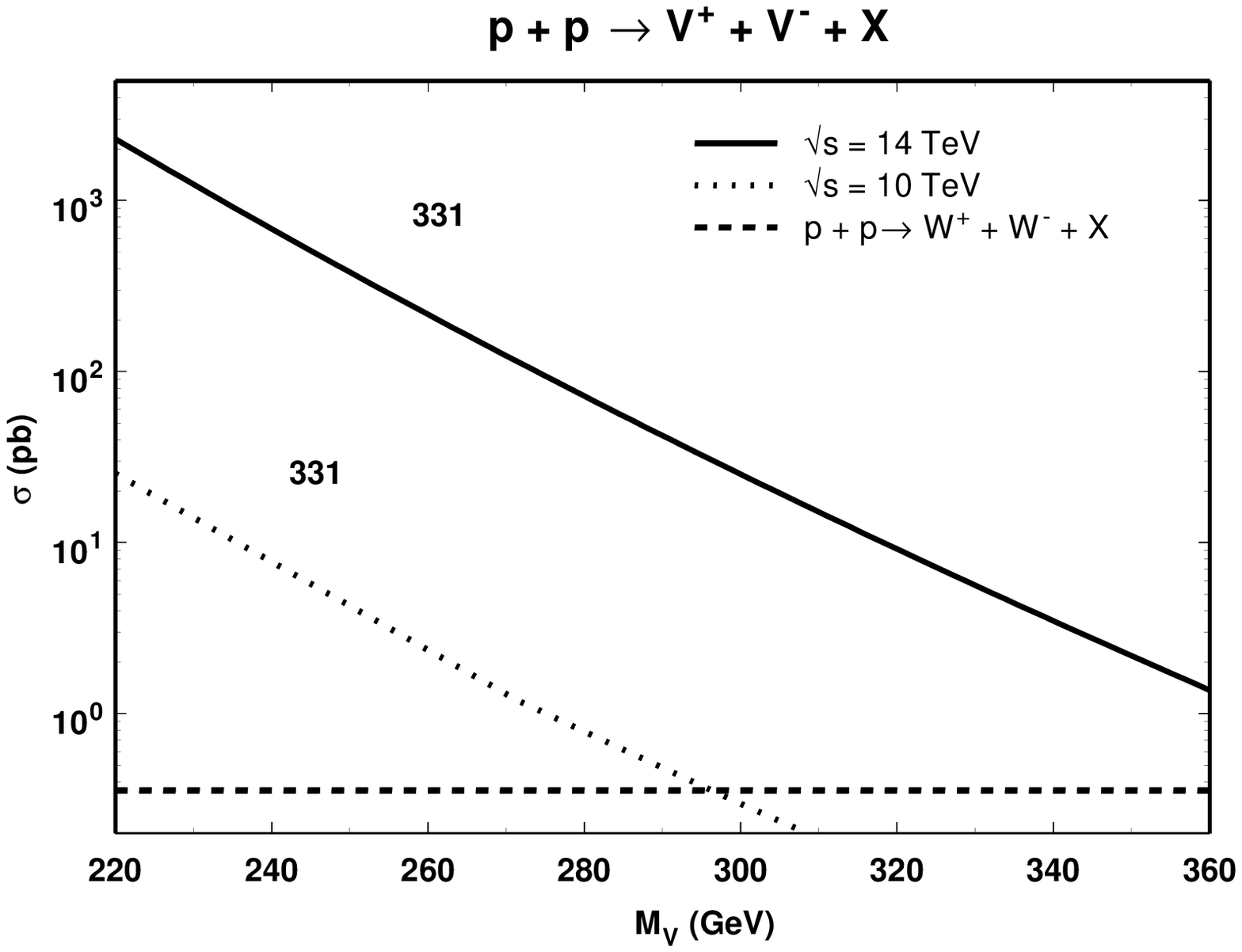}
\caption{The total cross section for the process $ p + p \lra V^+ + V^- + X$ against $M_V$ for the 3-3-1 model considering  $\sqrt s= 10$ TeV and  $\sqrt s= 14$ TeV. The horizontal line is the SM.}
\end{figure}

\par
\vskip 1cm

\section{Conclusions}
In this paper we focus on the bilepton ($V^{\pm}$) pair production in $pp$ collision at LHC. This particle is predicted in many extensions of the SM and in particular in the 3-3-1 model used in the present paper. We restrict our calculation to a version of the model where the bilepton mass is related to the mass of the extra neutral gauge boson $Z^\prime$, also predicted in the model. For the range of the extra neutral gauge  boson mass considered here, the dominant $V^{\pm}$ decay is leptonic ($\nu_\ell + \ell$). The  hadronic channel ($J_i + q_i$)  will open when $M_{Z^{\prime}}= 2 $ TeV, associated with a bilepton heavier than $600$ GeV. For the elementary Drell-Yan process, there are the contributions of $\gamma$, $Z$,  and $Z^\prime$ in the s-channel and the exotic quark  in the t-channel. $J_1$ is exchanged when the initial quark of colliding proton is an up-type quark, but when the down-type quark is participating,  $j_2$ and $j_3$ are exchanged. This is a consequence of our choice for family quark representation.

The correct high energy behavior of the elementary cross section follows from the balance between the individual contributions. In order to emphasize the role of the exotic quark contribution, we present in the Figure 5 the "bad" behavior of the elementary cross section in the absence of the t-channel contribution for a fixed $Z^\prime$ mass, we 
see clearly  that  the $u \bar u$ sub-process violates "softly" the unitarity bound and 
$d \bar d$  leads to a more severe violation. As expected, the $d \bar d$ process is more sensitive than $u \bar u$, because this channel receives additional exotic quark contribution. 

When considering the t-channel contribution  we take into account the mixing of quark mass eigenstates originated  from the Yukawa coupling. In this work we have obtained a set of mixing parameters allowing to a good behavior for the elementary cross section, for different $M_{Z^\prime}$. These parameters are related to our particular choice for $SU(3)_L$  family representation. This result  does not exclude any other choice for quark representation.

 In the  3-3-1 model, $Z^\prime$ couples to the quarks in a non universal way leading to  the existence of flavor changing vertices. As a consequence, the quark mixing parameters are also present in $Z^\prime$ quark  vertices. We display in the  Tables I and II our results for the flavor changing couplings.

In order to obtain the total cross section for the production of bilepton pairs we employed the  cut on the final particle pseudo-rapidity. Considering a  conservative integrated luminosity value and $\sqrt s = 10$ TeV we predict the production of a thousand of  $M_V \simeq 300$ GeV pairs mainly due to the contribution from $M_{Z^\prime}\simeq 1 $ TeV. For $\sqrt s = 14$ TeV the same number of events is obtained for  $M_V \, \simeq 450$ GeV pairs, associated to a $M_{Z^\prime}\simeq 1.4 $ TeV. One can ask about the possibility of Tevatron to find a large amount of bileptons. In fact, as our prediction lies on a large $M_{Z^\prime}$ (greater than $800$ GeV),  the required energy {\it per} quark for an individual  sub-process $q \bar q$ would be larger than  $500$ GeV not available at  Tevatron, where the  energy beam is about $900$ GeV. For this reason the Tevatron gives $M_{Z^\prime} > 600$ GeV.  

Finally, we observe that it is possible to distinguish the  leptons coming from bileptons  with those from the background of $W$ decay. In contrast with $W^\pm$ which decays into  $\bar \nu_\ell$ $\ell$,  the charged lepton coming from the bilepton decay has a large  transverse momentum. A useful  $p_T$ cut can eliminate this SM background.

We conclude that a large number of single charged bilepton pairs can be produced in the early stage of the LHC.

\vskip 1cm
\textit{Acknowledgments:} 
We thank Prof. V. Pleitez and Prof. F. Schwab for useful discussions. E. Ramirez Barreto thanks Capes and Y. A. Coutinho  thanks FAPERJ for financial support.

\end{document}